# Localizing Individual Exciton on a Quantum Hall Antidot

Rui Pu[1], Naomi Mizuno[1], Fernando Camino[2], Runchen Li[1], Kenji Watanabe[3], Takashi Taniguchi[4], Dmitri Averin[1]* and Xu Du[1]*

[1]*Department of Physics and Astronomy, Stony Brook University, Stony Brook, New York 11794-3800, USA*
[2]*Center for Functional Nanomaterials, Brookhaven National Laboratory, Upton New York 11973, USA*
[3] *Research Center for Electronic and Optical Materials, National Institute for Materials Science, 1-1 Namiki, Tsukuba 305-0044, Japan*
[4] *Research Center for Materials Nanoarchitectonics, National Institute for Materials Science, 1-1 Namiki, Tsukuba 305-0044, Japan*

**Abstract**

Quantum Hall systems host quasiparticles demonstrating correlated electron physics and non-trivial quantum statistics. Excitonic phases, archetypical for interaction effect, have attracted significant interest in recent years in double-layer quantum Hall systems where spatially separated electrons and holes form bosonic condensate through Coulomb interaction. Here, employing the approach of quantum Hall antidot with two spatially separated edge channels, we demonstrate a new type of quantum Hall quasiparticle exciton which represents a quantum-coherent bound state of an electron and a hole situated on their corresponding edges coupled through intralayer tunneling and Coulomb interaction. Quantum-coherent dynamics of the exciton is reflected in the observed evolution of the position and magnitude of the antidot conductance peaks around the electron-hole resonance. The quantum Hall antidot setup allows localization and electrical tuning of individual quantum Hall excitons. Quantum superposition of vacuum- and electron-hole pairing states is observed through the gate-dependent tunneling conductance of the antidot. Modeling the electron-hole pair as a coupled two-level system, semi-quantitative understanding of experimental observations is achieved. This work opens avenues for creating quantum systems of multiple quantum Hall quasiparticles.

*Corresponding authors: xu.du@stonybrook.edu, dmitri.averin@stonybrook.edu

Solid-state systems have been a major area for the development of quantum devices and circuits owing to their advantages in scalability, electric control, and constant progress in size reduction down to the nanoscale level. This combination of characteristics allows manipulation of individual quasiparticles associated with the states of charge[1-8], spin [9-14], magnetic flux and superconducting phase [15-21], or more complex combined degrees of freedom. In this context, two-dimensional (2D) electron quantum Hall (QH) system has been a unique platform which, in addition to the basic integer charge states, hosts the more exotic quasiparticles associated with fractional QH (FQH) abelian and non-abelian anyonic states [22-26], and excitons in double-layer QH systems [27-32] which exhibit Bose-Einstein condensation (see, e.g., Ref. [33]) and unusual quantum solids [34]. Evidence of the anyonic exchange statistics was obtained in the edge modes-based Fabry-Perot QH interferometer [35-39] and in the anyon collisions [40].

Another experimental platform which utilizes QH edge modes is QH antidot [41-43] in which the edge modes encircle the maximum of an electrostatic potential, forming closed orbits. Quantized energy levels obtained in this way can localize individual QH quasiparticles, including those with fractional charge and statistics, and provide the basis for their manipulations. Coulomb blockade-type correlations among the quasiparticles on the antidot, combined with the discrete antidot energy spectrum control the number of quasiparticles and minimize the phase space for scattering hence enhancing quantum coherence. Resonant tunneling through the antidot energy levels has successfully demonstrated the fractional charges of FQH quasiparticles [44-46]. The fractional exchange statistics of the anyonic quasiparticles is also theorized to be observable through transport measurement in multi-antidot "molecule" systems [47]. Beyond single antidot "atoms", however, while there has been experimental evidence of the double-antidot transport in accidental [48] and defined [42] structures, creating controllable and quantum coherent multiple QH quasiparticles systems remains limited and challenging.

As a step in this direction, here we demonstrate a coherent and interacting quantum state of individual localized exciton, achieved in a QH antidot platform. Figure 1(a) illustrates the basic principle of such QH antidot exciton in comparison with conventional exciton. A conventional exciton forms in the free space of a semiconductor by exciting a valence electron. The resulting hole in the valence band and the excited electron forms an excitonic pair through Coulomb interaction, with energy lower than the conduction band edge. A QH antidot exciton, on the other hand, forms in a pair of closely placed QH antidot edge modes hosting electron and hole. The quantum tunneling of an electron from the hole edge mode to the electron edge mode creates a coherent electron-hole pair which, through Coulomb interaction, forms an excitonic state with reduced ground state energy. In contrast to the QH excitons on bulk double-layer system which induces bosonic condensate through interlayer Coulomb interactions (Figure 1(b)) [28-32], an QH antidot exciton is localized individually and created via *intralayer* Coulomb interaction and tunneling.

While Coulomb blockade transport of individual excitons has been studied in quantum dots [49-51] as one of the approaches to the development of ``on-demand'' single-photon sources, and continue to attract practical interest up to now (see, e.g., Ref. [52] and references therein), the novel element added by our antidot structure to the manipulation of the individual excitons is quantum coherence between the electron and the hole parts of the exciton. Such quantum coherence within the individual excitons is potentially useful for quantum information applications, e.g. as an interface for quantum information transfer between the photons and the antidot qubits. From the point of view of qubits, electron-hole attraction within the exciton enhances the energy gap between the qubit states beyond the typically not-so-large tunnel coupling of these states, something that would be beneficial in many different contexts, most directly, in the case of the ground-state quantum computing [53].

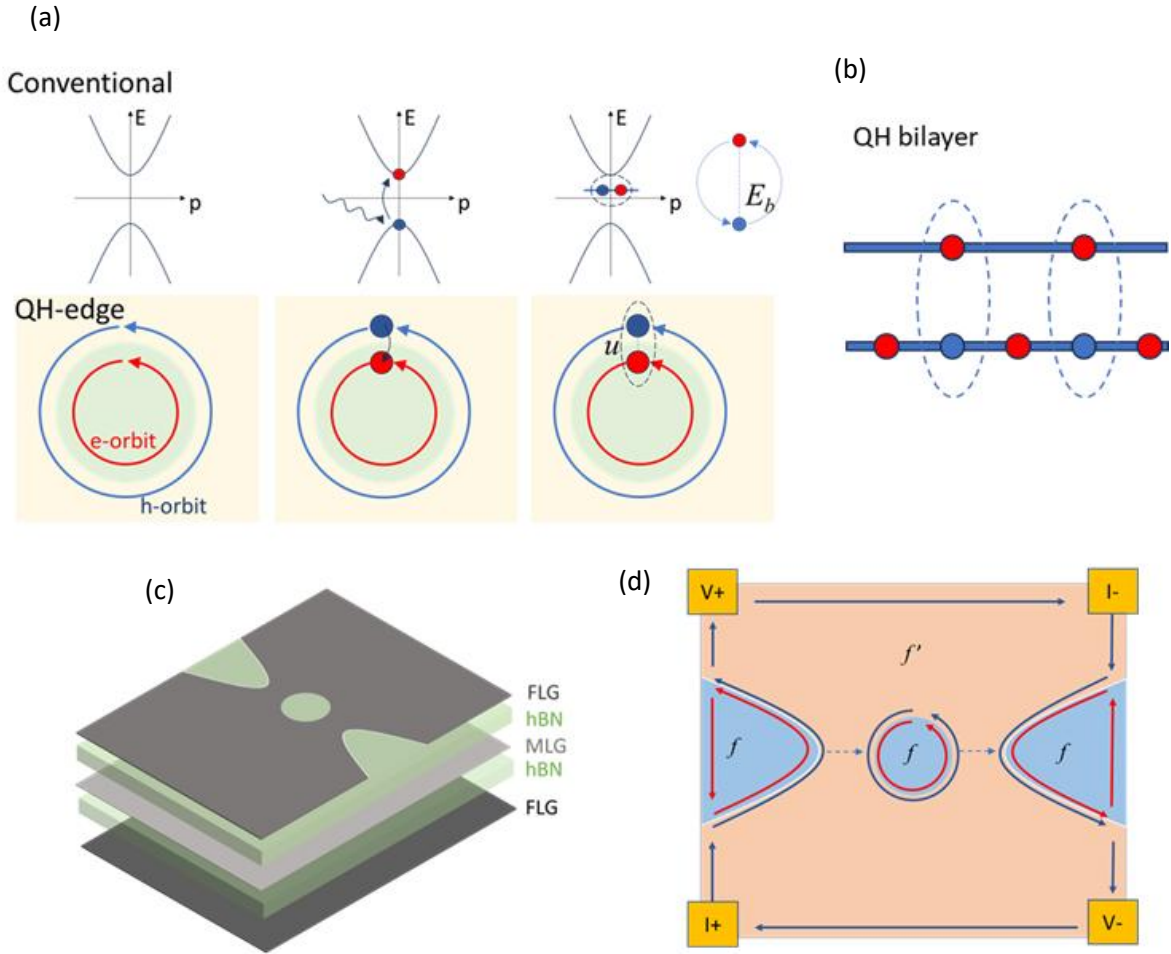

*Figure 1. The basic principle of QH exciton and experimental scheme. (a) Schematic showing principle of QH edge mode exciton formed by intralayer Coulomb interaction, and its comparison with conventional exciton. (b) Schematic showing the principle of QH bilayer exciton, formed by inter-layer Coulomb interaction. (c) Schematic of the sample consists of a heterostructure of patterned few-layer graphite (FLG) top gate, hBN, monolayer graphene (MLG), hBN and FLG bottom gate. (d) QH edge modes on a graphene channel, including the bulk edge modes and the top gate-defined edge modes at the couplers and the antidot. The edge modes form at the boundaries of different filling factors. The dotted arrows indicate where tunneling can happen.*

The experimental platform studied in this work is a graphene-hexagonal boron nitride (hBN) heterostructure illustrated in Figure 1(c), with hBN encapsulated monolayer graphene tunable by two graphite gates (see Supplementary Information (SI)). The bottom graphite gate tunes the carrier density uniformly throughout the whole monolayer graphene channel. The top graphite gate is patterned with an antidot and two "couplers" etched out. The diameter of the antidot here is $D \simeq 180$nm. The couplers point toward the antidot with an edge-edge distance of ~80nm at the closest points. Combining top and the bottom gates, carrier density and QH filling factor can be separately adjusted inside the antidot and coupler region, $\nu = \frac{nh}{eB}$, and outside, $\nu' = \frac{n'h}{eB}$. Here $n$ and $n'$ are the corresponding local carrier densities, $h$ is the Planck's constant, $e$ is the electron charge and $B$ is the external magnetic field. When the corresponding numbers of the filled Landau levels ($f$ and $f'$, positive on the electron side and negative on the hole side) are different between the two regions, QH edge modes form at the boundaries of the antidot and the couplers. As a result of quantum confinement ("particle in a ring") the energy of the QH edge modes encircling the antidot becomes quantized:

$$\varepsilon_j = \frac{2\hbar v}{D_{QH}}(j + \frac{\Phi}{\Phi_0}) \quad (1)$$

with energy level spacing: $\Delta\varepsilon = \frac{2\hbar v_{QH}}{D_{QH}}$ in fixed magnetic field [43, 46] (see SI). Here $j$ is an integer which indexes the quantization energy levels, $v_{QH}$ is the QH edge mode velocity, $D_{QH}$ is the effective diameter of the antidot edge mode, $\Phi$ is the magnetic flux through the antidot and $\Phi_0 = \frac{h}{e}$ is the magnetic flux quantum. The occupation of the energy levels can be tuned through the gate voltage; while a magnetic flux shifts the manifold of the levels, manifesting the Aharonov-Bohm effect. When the distance between the couplers' and the antidot's edge modes is sufficiently small, coupling between the bulk edge modes through the antidot takes place and lowers the transverse resistance $R_T \equiv \frac{V^+ - V^-}{I^+ - I^-}$ which can be characterized through the transport measurement in the configuration shown in Figure 1(d). In the regime when the bulk graphene is tuned to a QH plateau with quantized resistance $R_{QH}$, the antidot tunneling conductance can be measured from the corresponding resistance valley: $G_{AD} = \frac{1}{R_T} - \frac{1}{R_{QH}}$.

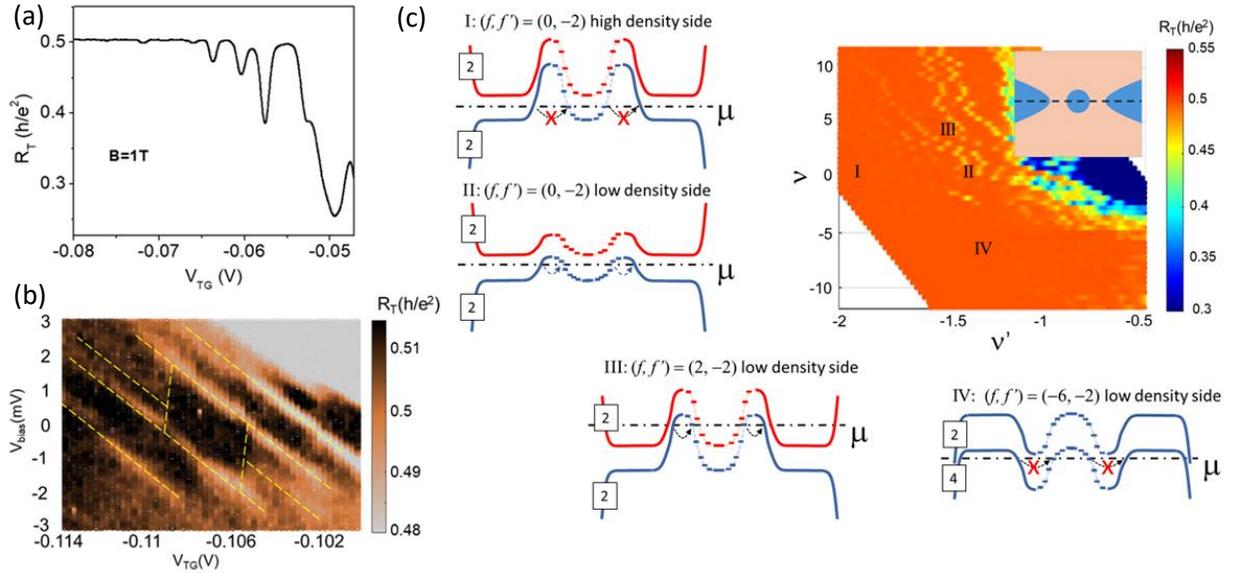

*Figure 2. Transport characteristics and tuning of QH antidot. (a) Resistance oscillations on the $f' = -2$ plateau from antidot tunneling, under zero back gate voltage. (b) Charge stability diagram for antidot tunneling on the $f' = -2$ plateau, showing the bias voltage ($V_{bias}$) and $V_{TG}$ dependence of the differential resistance. The dotted lines are guides to the eyes to some of the conductance peaks. Here back gate voltage is $V_{BG} = 0.08V$. (c)Tuning antidot tunneling through the dual gates. The color map shows $\nu'$ dependence of transverse resistance on the lower density side of the $f' = -2$ plateau over a wide range of $\nu$. The bright-colored dots correspond to the resistance valleys on a background of the $f' = -2$ plateau. Four characteristic regimes are labeled, with their corresponding Landau level energy diagrams illustrated. The Landau level diagrams are across center of the antidot and couplers, reaching the bulk*

*edges of the graphene channel, as indicted by the dotted line on the device schematic in the color map inset. The number on each Landau level indicates its degeneracy. Around the antidot, the Landau level energy is quantized into discrete energy levels due to quantum confinement effect on the closed edge modes. In regimes I and IV, antidot tunneling is suppressed by the large tunneling distances between the edges. Regimes II and III allow antidot tunneling.*

The basic signature of QH antidot tunneling is illustrated in Figure 2(a) under the simplest configuration where the antidot and the couplers are within the $f = 0$ insulating gap, while the outside bulk is on the $f' = -2$ QH plateau. Here, consistent with the prior studies on physically etched QH antidots [43, 46], the emergence of quasi-periodic resistance valleys on a QH plateau, associated with tunneling through discrete antidot levels, is observed in our gate-defined QH antidot device. Each resistance valley corresponds to adding a single hole charge onto the antidot [43]. Figure 2(b) goes beyond the linear conductance and demonstrates the top gate voltage and bias voltage dependencies of resistance ("charge stability diagram"). The dotted lines highlight the discernible Coulomb diamonds and excited states. We note that compared to typical quantum dots, the Coulomb diamond features are less pronounced in the QH antidot. This may be due to 1) break-down of the quantized plateau resistance under finite bias voltages which adds a bias-dependent background; 2) small Coulomb energy due to the relatively large antidot size and screening from the hBN dielectric layers; 3) asymmetric tunneling amplitude between the antidot and the two coupler, which results in nearly invisible resonant tunneling traces at the positive-sloped boundaries of the Coulomb diamonds. From this diagram, the on-site Coulomb repulsion energy and quantization energy spacing on a $f' = -2$ QH plateau in 1T magnetic field can be estimated to be $U \simeq 1$meV and $\Delta\varepsilon \simeq 0.7$meV, respectively. More examples of charge stability diagrams and Coulomb diamonds showing similar energy scale can be found in the SI. Based on the quantization energy, QH edge mode velocity of $v_{QH} = \frac{\Delta\varepsilon D_{QH}}{2\hbar} \simeq 10^5$m/s can be estimated, which is in qualitative agreement with the previous reports on edge mode velocity in QH interferometers[54, 55]. The measured on-site Coulomb energy of one edge state can be understood considering a localized ring of charge density with thickness on the order of the magnetic length $l_B = \sqrt{\frac{\hbar}{eB}}$, and partially screened by the nearby gate electrodes: $U = \frac{e^2}{2\pi\epsilon_{BN}\epsilon_0 L} \ln\left[\left(1 + \frac{4t_{BN}^2}{l_B^2}\right)^{1/2} + \frac{2t_{BN}}{l_B}\right]$ (see SI). Here $\epsilon_{BN} \simeq 3.5 \pm 0.2$ is the dielectric constant of hBN, and $t_{BN} \simeq 10$nm is the effective hBN thickness for the screening effect from both graphite gates. The observed Coulomb energy of $U \simeq 1$meV suggests an antidot mode circumference of $L \simeq 600$nm, i.e., $D_{QH} \simeq 190 \pm 10$nm. This is consistent with the similar but slightly smaller diameter of the antidot in the top graphite gate ($D \simeq 180$nm).

The gate-control of the antidot tunneling is illustrated in Figure 2(c). The main panel shows the tunneling resistance versus the filling inside ($v$) and outside ($v'$) the antidot/couplers regions. Sharp bright dots (resistance valleys) signify the antidot tunneling on the QH plateau of $R_{QH} = \frac{h}{2e^2}$ ($f' = -2$), coarsely sampled over a wide range of filling factor $v$ in the antidot/couplers regions that are tuned by the back gate. Four characteristic $(f, f')$ combinations are highlighted with their corresponding Landau level energy diagrams. The spatial profiles of the Landau level energy (in relation to the chemical potential) are tuned through the patterned top gate, while the back gate uniformly shifts the chemical potential throughout the whole channel. The QH edge modes are located where the Landau levels cross the chemical potential $\mu$. The edge mode tunneling strength depends sensitively on the spatial separation between the coupler and antidot edge modes, adjusted by the dual-gates. To enable coupler-antidot tunneling, the tunneling distance between the coupler edge mode and the antidot edge mode needs to be near its minimum, when Landau Level associated with these edge modes is tuned slightly across the chemical potential. Directly relevant to the discussions below and for $ff' \leq 0$, this corresponds to the lower carrier density end of a QH plateau. A more detailed discussion of the edge mode tunneling in these four regimes is given in the SI.

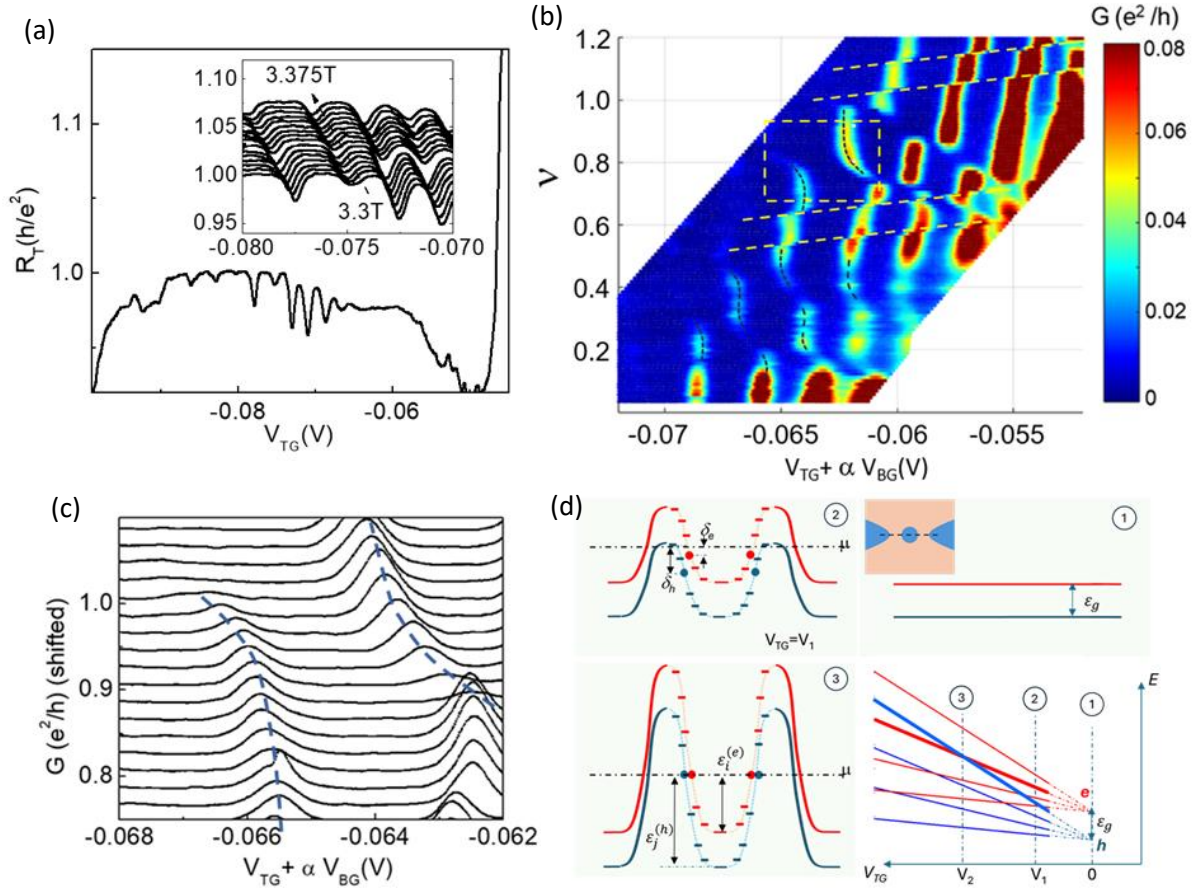

Figure 3. Localized QH exciton on an antidot. (a) Top gate dependent transverse resistance showing quasi-periodic valleys on the $f' = -1$ plateau. Inset: resistance oscillations showing shift in their gate voltage dependence under varying magnetic fields. The curves are displaced in y-axis for clarity. (b) Color-mapped antidot tunneling conductance versus doping inside and outside the antidot. The yellow dotted straight lines trace the random jumps in conductance peaks traces. The slope of these lines is identified to match with having constant top gate voltages. The black dotted curves are guides to the eyes for the anti-crossing-like features which are associated with tunneling through the electron-hole pairing state. The yellow dotted rectangle highlights a pronounced exciton state which is discussed quantitatively in the text. (c) Antidot tunneling conductance curves inside the yellow dotted rectangle in (b), shifted in the y-axis direction for clarity. The dashed lines highlight the evolution of the position of the conductance peaks that represent the electron-hole resonance. Note that the system of peaks in the lower right corner of the plot not marked by the dashed line belongs to the next quantization energy level and are not related to the electron-hole resonance. (d) Qualitative schematic showing gate tuning of the antidot levels for the outer (hole) and inner (electron) antidot modes. Bottom right plot: the energy levels spread out with increasing magnitude of top gate voltage. Due to the energy offset between the $f' = \pm 1$ Landau levels

($\varepsilon_g$), the electron- and hole- antidot levels can cross, resulting in resonances and pairing. Here we highlight the shift of an electron level and a hole level (thick red and blue lines in the $E(V_{TG})$ plot and, the red and blue circles in the Landau level diagrams) under changing top gate voltage, with corresponding Landau level diagram at top gate voltages of 0, $V_1$ and $V_2$ shown in panels ①, ② and ③. Panel ① inset: device schematic where Landau level diagrams are along the dotted line.*

Having demonstrated the basic principle of QH antidot operation, we next focus on the symmetry-breaking $f' = -1$ state in a magnetic field $B = 3.3$T. Figure 3(a) shows the top gate voltage dependence of the transverse resistance $R_T$, where quasi-periodic resistance valleys associated with coupler-antidot tunneling of the hole side of the zeroth Landau level edge modes emerge on the plateau background of $R_{QH} = \frac{h}{e^2}$. Following eq.(1) the gate voltage positions of these valleys shift with changing magnetic field at a rate of one resistance valley per magnetic flux quantum $\Phi_0$ through the effective area of the antidot: $\frac{\delta V_{TG}}{\Delta V_{TG}} = -\frac{\delta \Phi}{\Phi_0} = -\frac{\delta B \pi D_{QH}^2}{4\Phi_0}$. Here $\delta V_{TG}$ is the shift of a resistance valley in top gate voltage under a changing magnetic field, $\Delta V_{TG}$ is the top gate voltage period of the resistance valleys, $\delta B$ and $\delta \Phi$ are the changes of magnetic field and flux, respectively. From the magnetic shift of the resistance valleys, $D_{QH} \approx 230nm$ can be estimated suggesting the edge mode to situate ~25nm away from the physical antidot edge of the top graphite gate. With this effective diameter we can also estimate the number of charges added to the antidot per gate period: $\Delta N = \frac{c_{TG} \pi D_{QH}^2}{4e} \Delta V_{TG} = 1.04 \pm 0.06$ ($c_{TG} = \frac{\epsilon_0 \epsilon_{BN}}{d_{hBN}}$ is the area capacitance of the top gate with $d_{hBN} = 18nm$ being the thickness of the hBN layer between graphene can the top graphite gate), consistent of addition of single fundamental charges in the integer QH regime[46]. All these observations further confirm the antidot nature of the resistance valleys.

The main new finding of this work is shown in Figure 3(c) which illustrates the evolution of the antidot tunneling conductance oscillations on the lower density side of the $f' = -1$ plateau over a range of filling factor $\nu$ in the antidot/coupler regions. For clarity, the horizontal axis is set to $V_{TG} + \alpha_{AD} V_{BG}$, which depicts directly the electrostatic doping at the outside vicinity of the antidot, where $\alpha_{AD} = 0.53$ is the effective capacitance ratio between the bottom and the top gates for antidot charging (see SI). The antidot tunneling conductance peaks appear as near-vertical stripes with abrupt jumps. Most interestingly, anti-crossing-like features are observed, highlighted by the black dotted curves in Figure 3(b). These features are generally present under the $(f, f') = (1, -1)$ configuration, although the regularity of their locations is disturbed by a few random slides occurring along the yellow dotted straight lines in Figure 3(b). The source of these random slides is identified as quantum dot-like charge trapping centers located near the antidot edge on the top graphite gate (see SI), which is not the focus of this work. In the discussions below we focus on the anti-crossing feature highlighted by the yellow box in Figure 3(b), for which the individual curves of the antidot conductance are shown in Figure 3(c). As will be argued below after developing a quantitative model, the anti-crossing-like features observed here originate not from atomic-scale traps, but from the formation of a gate-defined pair of electron-hole edges coupled both by tunneling amplitude and Coulomb attraction.

Anti-crossing features are typically associated with level coupling. In high quality graphene samples, symmetry-breaking interaction effects lift the 4-fold degeneracy of the zeroth Landau level. With well-developed $f' = -1$ plateau and insulating behavior at the charge neutrality, the inner electron and the outer hole edge modes are expected to coexist and be spatially separated by a gapped region. Increasing the magnitude of the top gate voltage, the antidot-defining potential profile becomes increasingly sharp, resulting in antidot edge modes with larger edge mode velocities ($v_{QH} \sim \frac{dV}{dr}$) hence larger quantization energy level spacing ($\Delta\varepsilon = \frac{2\hbar v_{QH}}{D_{QH}}$). As a result, both the outer hole- and the inner electron- antidot edge modes have their quantization energy levels "fan out" with increasing top gate voltage. Since the manifolds of hole- and the electron- quantization energy levels are offset by the charge neutrality gap, the energy levels of the outer- and inner- modes with different indices will have different shifting rates under changing top gate voltage, and will cross each other as illustrated by Figure 3(d). As a result, just from the top gate voltage, some of the hole- and electron levels can be tuned to align with and cross each other, creating tuning and detuning of an electron-hole resonance. With the back gate tuning the chemical potential throughout the channel, one can ensure tunneling through the antidot under the conditions of the electron-hole resonance. A more analytical discussion of the gate voltage dependence of the resonance can be found in the SI.

To understand the energy spectrum of the coupled electron and hole edge modes, we consider a basic model of graphene QH antidot[43] extended to two coupled co-centric edge modes, shown in Figure 4(a). Here $\Gamma$ is the tunneling rate between the couplers and the outer hole mode. The inner electron mode is not coupled directly to the couplers, but to the hole mode with a tunneling amplitude $\Delta$. The energy spectra of the two edge modes are quantized, and therefore, in the main approximation, the tunnel coupling between the two modes is coherent. Besides tunneling, electrons and holes also experience Coulomb interaction, which can be described in the constant-interaction approximation for each individual mode:

$$U(n_h, n_e) = \frac{1}{2} U_h n_h (n_h - 1) + \frac{1}{2} U_e n_e (n_e - 1) - u n_h n_e \quad (2)$$

Here $n_h$ and $n_e$ are the numbers of particles in the hole and electron mode respectively, $U_{h,e}$ are the corresponding interaction constants for the two modes, and $u > 0$ is the magnitude of the Coulomb electron-hole attraction between the two edges.

Similar to the basic Coulomb-blockade model for tunneling through the quantum dots, Eq.(2) leads to periodic response of the edges to the two gate voltages that control $n_h$ and $n_e$. For the outer edge, this translates into a quasi-periodic sequence of the resonant tunneling peaks of the antidot conductance $G_{AD} = dI/dV_{bias}|_{V_{bias}=0}$, with each peak occurring at the bias conditions at which the number of holes in the edge changes, $n_h \to n_h + 1$. One the other hand, the relations among the parameters in Eq. (2): nearly equal repulsion energies $U_e$ and $U_h$, and strong attraction constant $u$ close to $U$, combined with the choice of the non-vanishing tunneling terms, make this model very specific to the system we consider: concentric electron-hole edges encircling one antidot. In this case, the inner edge does not affect the conductance directly, but the number $n_e$ of electrons in it also changes periodically by 1 as a sequence of resonances.

To describe the antidot transport properties in the vicinity of one combined electron-hole resonance (where both the electron and hole antidot energy levels align with the chemical potential), we need to consider four charge states, each labeled by $n_h$ and $n_e$: $|n_h n_e\rangle$. On the basis $\{|00\rangle, |11\rangle, |01\rangle, |10\rangle\}$ of these four states the Hamiltonian of the antidot is:

$$H = \begin{pmatrix} 0 & \Delta^* & 0 & t^+ \\ \Delta & \xi & t & 0 \\ 0 & t^+ & -\delta_e & 0 \\ t & 0 & 0 & \delta_h \end{pmatrix} \quad (3)$$

Here $\delta_h$ is the gate-induced energy deviation of the outer edge state from the resonance point, $\delta_e$ is the same for the inner edge, $\xi = \delta_h - \delta_e - u$ is the energy of the electron-hole pair occupying the edge states. We include $t, t^+$ terms to indicate how the states of the electron-hole system are coupled to the external charge reservoirs and produce the incoherent tunneling rates $\Gamma$ illustrated in Figure 4(a). However, under the limit of weak antidot-reservoir coupling we do not calculate these terms. Note that the electron-hole tunnel amplitude $\Delta$ that couples $|00\rangle$ and $|11\rangle$ states has a somewhat unusual form of simultaneously creating or destroying two particles, electron and hole which form an exciton. The reason for this is that it describes the coupling of the two different edges states, the electron and the hole edge states. If one reformulates the description of both edges in terms of the same-charge particles, e.g., electrons, it becomes evident that the tunnel amplitude $\Delta$ corresponds physically to the usual tunneling process of transfer of one electron between the two edge states. Such a transfer either adds electron to the electron edge leaving a hole in the hole edge thus creating an exciton, or, in the reverse process, takes electron from the electron edge filling a hole in the hole edge, thus destroying an exciton in the process.

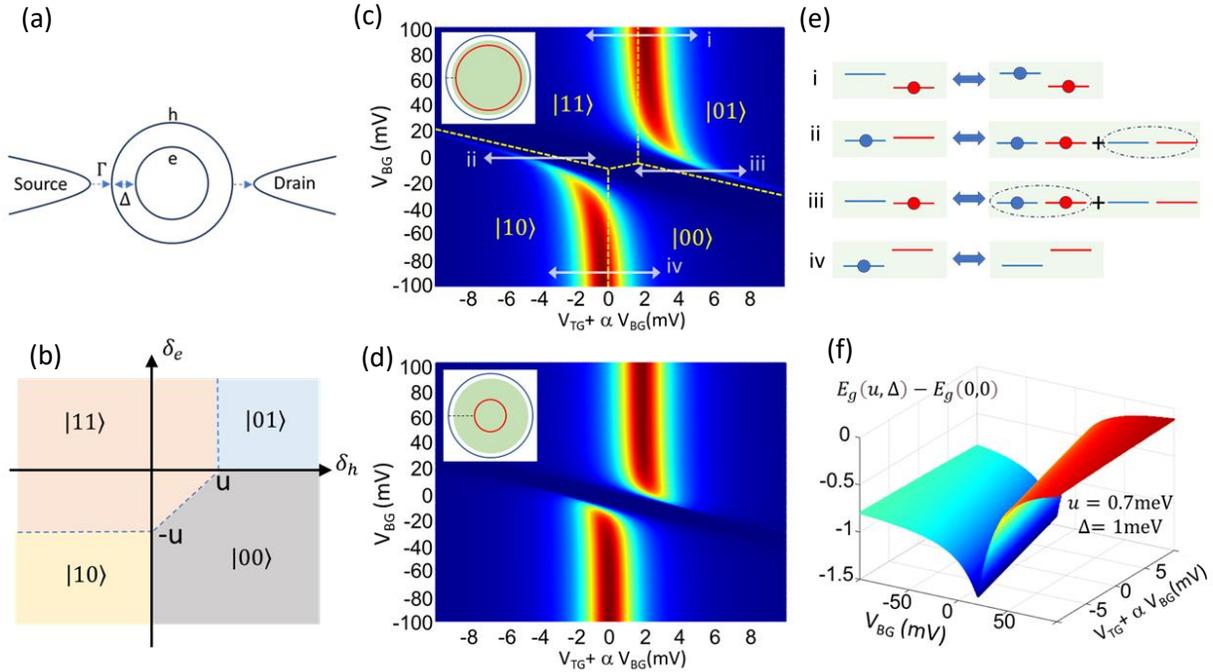

Figure 4. Modeling the coherent electron-hole pair state. (a) Transport model of the antidot made of two coupled electron and hole edge modes. (b) Diagram of charging phases (controlled by the two gate voltages) with different electron-hole configurations as the ground states. The condition $\xi = 0$, where $\xi$ is the energy difference between the states $|00\rangle$ and $|11\rangle$ introduced in Hamiltonian (3), is shown by the dashed line. The states $|00\rangle$ and $|11\rangle$ are degenerate along this line and therefore are mixed strongly by the tunnel amplitude $\Delta$. Without this mixing the conductance peaks due to tunneling through the outer edge mode happen directly along the boundary between the states $|11\rangle$ and $|01\rangle$ (described by the conditions $\delta_h = u$; $\delta_e > 0$) and between the states $|10\rangle$ and $|00\rangle$ ($\delta_h = 0$; $\delta_e < -u$). (c) and (d) illustrate color-coded tunneling conductance versus top and back gate voltages, computed with the transport model with $u = 0.7$meV and $\Delta = 1$meV, and $u = 0.7$meV and $\Delta = 0.4$meV, respectively. Correspondingly the insets illustrate the real space separation of the two modes under these parameters. In (c) the labeled gray arrows indicate tunneling processes illustrated in (e). Processes i and iv are between states with electron and hole levels detuned. Processes ii and iii involve tunneling through the excitonic state which is a superposition of $|00\rangle$ and $|11\rangle$, with dotted ovals indicate the components which contributes to the tunneling conductance. (f) Ground state energy reduction from that of the non-interacting isolated electron-hole pair.

From the Hamiltonian, $\delta_h$ and $\delta_e$- dependent ground states can be calculated as superpositions of the basis states. Tunneling through the antidot occurs when the system is tuned across two degenerate ground states with the same electron numbers but with hole numbers differing by one. This limits the conductance-contributing tunneling processes to $|00\rangle \leftrightarrow |10\rangle$ (empty electron edge mode) and $|11\rangle \leftrightarrow |01\rangle$ (filled electron edge mode). Without electron-hole coupling, the basis states are the eigenstates of the Coulomb energy, and their charge configurations in the ground state depend directly on the energies $\delta_h, \delta_e$ as shown in Figure 4(b). $|00\rangle \leftrightarrow |10\rangle$ and $|11\rangle \leftrightarrow |01\rangle$ lead to the vertical traces of conductance peak along the corresponding phase boundaries in the $(\delta_h, \delta_e)$ plane. The horizontal shift between the two vertical parts of this boundary (which is also observed experimentally, e.g., in Figure 3(c) is the direct manifestation of electron-hole attraction within the exciton. This shift represents the additional magnitude of the gate voltage that is needed to overcome the Coulomb attraction energy $u$ between the individual electron and hole in the two respective edge states of the antidot, after one extra electron was added to the internal edge of the antidot, when the gate voltages evolved past the resonance point.

Near electron-hole resonance, the tunnel amplitude $\Delta$ mixes the states $|00\rangle$ and $|11\rangle$ along $\xi = 0$, forming ground state which is the quantum superposition of the two states (see SI). Tunneling conductance peak emerges at energy determined by the condition of degeneracy between the resonance state and the adjacent ground states of $|01\rangle$ and $|10\rangle$ (see SI):

$$\delta_e = \begin{cases} \frac{|\Delta|^2}{\delta_h - u}, & \delta_h > u \\ \frac{|\Delta|^2}{\delta_h} - u, & \delta_h < 0 \end{cases} \quad (4)$$

Here the hyperbolic $(\delta_e, \delta_h)$ relations (which, in absence of electron-hole tunneling, correspond to the straight boundaries between the states $|11\rangle$ and $|01\rangle$ and between the states $|10\rangle$ and $|00\rangle$ in Figure 4b) give rise to the anti-crossing feature in the conductance peak trace.

Near the resonance, the amplitude of the tunneling conductance from the $|00\rangle \leftrightarrow |10\rangle$ and the $|11\rangle \leftrightarrow |01\rangle$ processes is determined by the composition of the $|00\rangle$ and $|11\rangle$ states in the superposition, respectively. The tunneling probability $p_0$ for the $|00\rangle \leftrightarrow |10\rangle$ process rapidly decays as the superposition evolves into pure $|11\rangle$ state under negative $\xi$; and the tunneling probability $p_1$ for the $|11\rangle \leftrightarrow |01\rangle$ process rapidly decays as the superposition evolves into pure $|00\rangle$ state under positive $\xi$ (see SI):

$$p_{0,1} = \begin{cases} \frac{1}{2}\left(1 + \frac{\xi/2}{\sqrt{\xi^2/4 + |\Delta|^2}}\right) \\ \frac{1}{2}\left(1 - \frac{\xi/2}{\sqrt{\xi^2/4 + |\Delta|^2}}\right) \end{cases} \quad (5)$$

Hence deviation from the resonance condition drives the system out of the electron-hole superposition, and results in the termination of the curvy anti-crossing feature

Applying the model above to our experimental system, we simulate the doping dependence of the antidot tunneling conductance as shown in Figure 4(c). Here we assume Lorentzian-shaped tunneling conductance peaks[46] with an energy broadening of 0.17meV to match with the broadening of the measured conductance peaks. The energy $(\delta_e, \delta_h)$ dependences of the peak position and peak amplitude are calculated from eqs.(4) and (5). The gate voltage-tuning of energy at ~0.5meV/mV is obtained from ratio between the top gate voltage period and the corresponding charging energy measured in Figure 2(a) and (b). Semi-quantitative agreement with the experimental observations (Figure 3b,c) is achieved with an electron-hole Coulomb energy of $u = 0.8$meV and tunneling amplitude of $\Delta = 0.3$meV, reproducing the anti-crossing-like gate-dependent peak position as well as the tunneling conductance peak amplitude. In Figure 4(c), the arrows highlight tunneling processes which give rise to conductance peaks, both far off-resonance (i and iv) and near resonance (ii and iii). The corresponding electron and hole energy level configurations of the tunneling states are depicted in Figure 4(e). In processes ii and iii, exciton states are the superposition of the $|00\rangle$ and $|11\rangle$ states.

For the Coulomb interaction between the electron and the hole QH antidot edges, we estimate $u = \frac{e^2}{2\pi\epsilon_{BN}\epsilon_0 L}\ln\left[\left(1 + \frac{4t_{BN}^2}{d_{eh}^2}\right)^{1/2} + \frac{2t_{BN}}{d_{eh}}\right]$ (see SI). Here $d_{eh}$ is the radial separation between the electron and hole edges which should roughly be equal to $d_{eh} \simeq 35$nm to produce $u = 0.8$meV, in agreement with observations. A theoretical estimation of $d_{eh}$ requires detailed information on the $\nu = 0$ energy gap and the gate-dependent edge mode velocity, and is beyond the scope of this study. Our QH antidot exciton estimates, nevertheless, are consistent with the most natural model of concentric electron and hole edge modes with similar diameters and significant tunneling between them. The order-of-magnitude of these estimates for $u$ and $\Delta$ are consistent with the assumed geometry of our exciton model with tunnel-coupled, roughly concentric electron and hole edges. Indeed, the inter-mode Coulomb attraction energy $u$ is close to and slightly smaller than the onsite Coulomb energy measured from the charge stability diagrams, consistent with the fact that the effective distance between the edges is slightly larger than the characteristic distance between the charges on the same edge. At the same time, the tunnel coupling $\Delta$ is significant which implies closely spaced electron and hole edges. If an electron would be localized on an atomic-scale impurity trap at such a small distance from the hole edge, it would produce more complex scattering of the hole instead of the observed coherent tunneling with amplitude $\Delta$.

Reducing the inter-mode coupling $\Delta$, the model suggests that the hyperbola part of the anti-crossing feature around the resonance is reduced (Figure 4d). This is consistent with the observations at smaller filling factor $\nu$ as shown in Figure 3b (more discussion can be found in the SI). At small $\nu$, the inner electron edge mode just starts to emerge near the center of the antidot and hence is physically distant from the outer hole edge mode. As a result, the tunneling amplitude between the two modes is small and the hyperbola feature in the gate dependence of the conductance peaks is largely reduced.

Similar to the conventional exciton, localized QH exciton state is stabilized by reducing the ground state energy of an electron-hole pair. Compared to the non-interacting ($u = 0$) and isolated ($\Delta = 0$) electron-hole states, the excitonic state has its ground state energy lowered by finite interaction and finite inter-mode tunneling: $E_g(u, \Delta) - E_g(0,0) < 0$. This gate-dependent energy reduction can be considered as the equivalent of exciton binding energy. In particular, for the neutrally biased electron-hole state ($\delta_h = \delta_e$), we have $E_g(0,0) - E_g(u, \Delta) = \frac{u}{2} + \sqrt{\frac{u^2}{4} + \Delta^2}$. Interestingly and different from conventional excitons, the binding energy of the localized QH exciton is not only affected by the Coulomb energy but also by the tunneling amplitude between the electron and the hole modes. Finite inter-mode tunneling leads to a ground state energy reduction in comparison to the decoupled interacting electron-hole pair: $E_g(u, \Delta) - E_g(u, 0) = \left|\frac{\xi}{2}\right| - \sqrt{\frac{\xi^2}{4} + \Delta^2}$, with maximum reduction of $\Delta$ occurring at $\xi = 0$ (see Figure 4(f)).

We note that a unique characteristic of the electron-hole system studied here is the combination of Coulomb repulsion on-site and Coulomb attraction between the two edges, which is different from the more familiar multi-quantum dot systems with Coulomb repulsion only. The electron-hole attraction $u$ in the exciton breaks one continuous trace of the external hole edge mode tunneling peak when the inner electron edge changes from being empty ($|10\rangle \leftrightarrow |00\rangle$) to filled ($|11\rangle \leftrightarrow |01\rangle$). Since the inter-edge Coulomb attraction effectively reduces the on-site Coulomb blockade, the shift of the off-resonance tunneling conductance peaks is towards smaller magnitude of gate voltages with increasing number of electrons on the inner antidot edge. Classically, the two segments of the conductance peak would be connected and smoothly shifted by an energy of $u$, reflecting the gradual increase of the average occupation of the electron state from 0 to 1 with the gate voltage. Quantum mechanically, on the other hand, strong tunneling couples the vacuum state $|00\rangle$ and electron-hole pair state $|11\rangle$, creating a coherent quantum superposition of the two states. Hole tunneling through the vacuum-pair superposition, as discussed through our model, leads to the anti-crossing-like gate voltage dependence of the conductance peaks with diminishing amplitude away from the electron-hole resonance.

In conclusion, we have demonstrated a localization of individual quantum-coherent QH exciton in a graphene QH antidot. Compared to the conventional excitons on semiconductors, the QH exciton allows electrical tuning of the electron and hole states, and shows a ground state energy which is determined both by Coulomb interaction and inter-mode tunneling. This work paves the way to a series of scientific tasks and technical developments. For example, excitonic states may be studied in more complex and more versatile many-antidot system for excitonic molecules; time-dependent measurements may be performed on the excitonic state to explicitly study the quantum coherence; excitonic states may also be formed with fractionally charge quasiparticles in the FQH regime, where anyon exchange and statistics may play a profound role in the physics properties of the excitons.

**Acknowledgement**

X.D. and D.A. acknowledges support from NSF awards under Grant No. DMR-2104781. K.W. and T.T. acknowledge support from the JSPS KAKENHI (Grant Numbers 21H05233 and 23H02052) , the CREST (JPMJCR24A5), JST and World Premier International Research Center Initiative (WPI), MEXT, Japan. This research used the Electron Microscopy facility of the Center for Functional Nanomaterials (CFN), which is a U.S. Department of Energy Office of Science User Facility, at Brookhaven National Laboratory, under Contract No. DESC0012704.